\documentclass[final]{ols}

\usepackage{url}
\usepackage{zrl}

\usepackage{framed}
\usepackage{color}
\definecolor{shadecolor}{rgb}{0.9,0.9,0.9}

\ifpdf
\usepackage[pdftex]{graphicx}
\else
\usepackage{graphicx}
\fi

\usepackage{multirow}

\begin{document}

% Mandatory: article title specification.
% Do not put line breaks or other clever formatting in \title or
% \shortauthor; these are moving arguments.

\title{Load-Balancing for Improving User Responsiveness on Multicore Embedded Systems}

% Subtitle is optional.
%\subtitle{extended thread api, thread naming \& profiling, thread scheduling framework }
\subtitle{2012 Linux Symposium}

% You can put a fixed date in if you wish,
% allow LaTeX to use the date of typesetting,
% or use \date{} to have no date at all.
% Whatever you do, there will not be a date
% shown in the proceedings.
%\date{Deadline: Jun-17-2012 19:00} 
%\date{Jun-17-2012 19:00} 
%\date{} 

\shortauthor{Geunsik Lim, Changwoo Min, YoungIk Eom}  % Just you and your coauthors' names.
% for example, \shortauthor{A.N.\ Author and A.\ Nother}
% or perchance \shortauthor{Smith, Jones, Black, White, Gray, \& Greene}

\author{%  Authors, affiliations, and email addresses go here, like this:
Geunsik Lim\\
{\itshape Sungkyungkwan University}\\
{\itshape Samsung Electronics}\\
{\ttfamily\normalsize leemgs@ece.skku.ac.kr}\\
{\ttfamily\normalsize geunsik.lim@samsung.com}\\
\and
Changwoo Min\\
{\itshape Sungkyungkwan University}\\
{\itshape Samsung Electronics}\\
{\ttfamily\normalsize multics69@ece.skku.ac.kr}\\
{\ttfamily\normalsize changwoo.min@samsung.com}\\
\and
YoungIk Eom \\
{\itshape Sungkyungkwan University}\\
{\ttfamily\normalsize yieom@ece.skku.ac.kr}\\
} % end author section

%\author{
%Matthew Dobson, Patricia Gaughen, Michael Hohnbaum \\
%{\em IBM LTC, Beaverton, Oregon, USA}\\
%{\tt\normalsize one@email.addr, two@email.addr, three@email.addr} \\

\maketitle

\begin{abstract}
% Article abstract goes here.
Most commercial embedded devices have been deployed with a single processor architecture. The code size and complexity of applications running on embedded devices are rapidly increasing due to the emergence of application business models such as Google Play Store and Apple App Store. As a result, a high-performance multicore CPUs have become a major trend in the embedded market as well as in the personal computer market. 

Due to this trend, many device manufacturers have been able to adopt more attractive user interfaces and high-performance applications for better user experiences on the multicore systems.

In this paper, we describe how to improve the real-time performance by reducing the user waiting time on multicore systems that use a partitioned per-CPU run queue scheduling technique. Rather than focusing on naive load-balancing scheme for equally balanced CPU usage, our approach tries to minimize the cost of task migration by considering the importance level of running tasks and to optimize per-CPU utilization on multicore embedded systems.

Consequently, our approach improves the real-time characteristics such as cache efficiency, user responsiveness, and latency. Experimental results under heavy background stress show that our approach reduces the average scheduling latency of an urgent task by 2.3 times.

\end{abstract}

% Body of your article goes here. You are mostly unrestricted in what
% LaTeX features you can use; however, the following will not work:
% \thispagestyle
% \marginpar
% table of contents
% list of figures / tables
% glossaries
% indices

% Article: 1. introduction.
\section{Introduction}

Performance improvement by increasing the clock speed of a single CPU results in a power consumption problems \cite{mem-aware-scheduling, thermal-aware-scheduling}. Multicore architecture has been widely used to resolve the power consumption problem as well as to improve performance \cite{cmp-aware-scheduler}. Even in embedded systems, the multicore architecture has many advantages over the single-core architecture \cite{embedded-multicore-processors}. 

Modern operating systems provide multicore aware infrastructure including SMP scheduler, synchronization \cite{using-os-observation}, interrupt load-balancer, affinity facilities \cite{improving-cpu-affinity, migration-policy-for-multicore}, CPUSETS \cite{cpusets}, and CPU isolation \cite{container-based-os-virtualization, shielded-cpus}. These functions help running tasks adapt to system characteristics very well by considering CPU utilization.

Due to technological changes in the embedded market, OS-level load-balancing techniques have been highlighted more recently in the multicore based embedded environment to achieve high-performance. As an example, the needs of real-time responsiveness characteristics \cite{rt-scheduling-on-multicore} have increased by adopting multicore architecture to execute CPU-intensive embedded applications within the desired time on embedded products such as a 3D DTV and a smart phone.

In embedded multicore systems, efficient load-balancing of CPU-intensive tasks is very important for achieving higher performance and reducing scheduling latency when many tasks running concurrently. Thus, it can be the competitive advantage and differentiation.

In this paper, we propose a new solution, \emph{operation zone based load-balancer}, to improve the real-time performance \cite{real-time-performance-and-middleware} on multicore systems. It reduces the user waiting time by using a partitioned scheduling---or per-CPU run-queue scheduling---technique. Our solution minimizes the cost of task migration \cite{impact-of-task-migration} by considering the importance level of running tasks and per-CPU utilization rather than focusing on naive CPU load-balancing for balanced CPU usage of tasks.

Finally, we introduce a flexible task migration method according to \emph{load-balancing operation zone}. Our method improves operating system characteristics such as cache efficiency, effective power consumption, user responsiveness, and latency by re-balancing the activities that try to move specific tasks to one of the CPUs on embedded devices. This approach is effective on the multicore-based embedded devices where user responsiveness is especially important from our experience.

% Article: 2. load-balancing mechanism on Linux.
\section{Load-balancing mechanism on Linux}

\begin{figure}
\centering
\includegraphics[width=1.0\columnwidth]{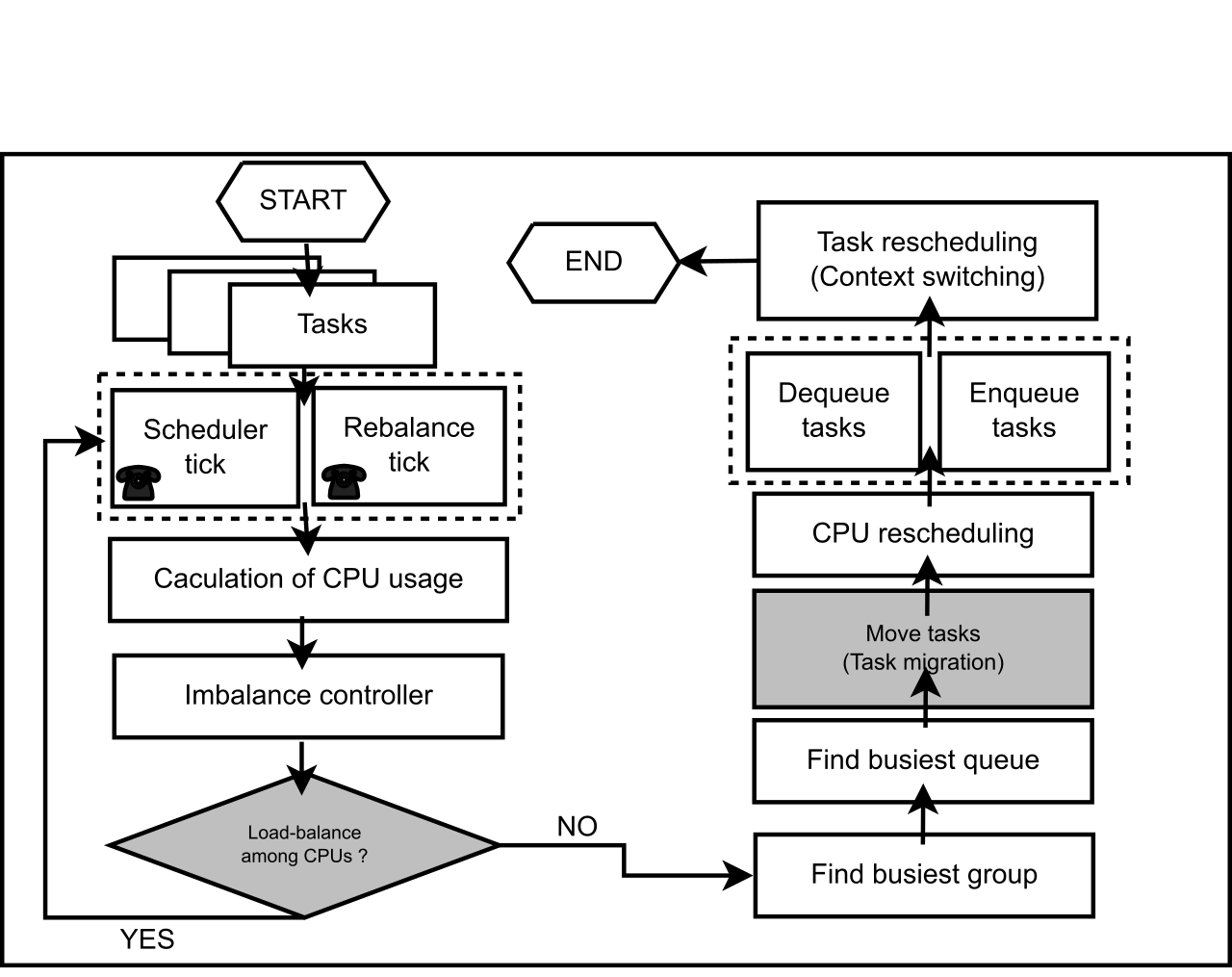}
\caption{Load-balancing operation on Linux}
\label{lim-fig-existing-loadbalancing}
\end{figure}

The current SMP scheduler in Linux kernel periodically executes the load-balancing operation to equally utilize each CPU core whenever load imbalance among CPU cores is detected. Such aggressive load-balancing operations incur unnecessary task migrations even when the CPU cores are not fully utilized, and thus, they incur additional cache invalidation, scheduling latency, and power consumption. If the load sharing of CPUs is not fair, the multicore scheduler \cite{understanding-linux-kernel-3rd} makes an effort to solve the system's load imbalance by entering the procedure for load-balancing \cite{exploiting-process-lifetime}. Figure~\ref{lim-fig-existing-loadbalancing} shows the overall operational flow when the SMP scheduler \cite{arm-smp-dma} performs the load-balancing. 

At every timer tick, the SMP scheduler determines whether it needs to start load-balancing \cite{improve-load-balancing} or not, based on the number of tasks in the per-CPU run-queue. At first, it calculates the average load of each CPU \cite{loosely-coupled-multicore}. If the load imbalance between CPUs is not fair, the load-balancer selects the task with the highest CPU load \cite{load-balancing-on-speed}, and then lets the migration thread move the task to the target CPU whose load is relatively low. Before migrating the task, the load-balancer checks whether the task can be instantly moved. If so, it acquires two locks, \ident{busiest->lock} and \ident{this_rq->lock}, for synchronization before moving the task. After the successful task migration, it releases the previously held double-locks \cite{task-migration-soc}. The definitions of the terms in Figure~\ref{lim-fig-existing-loadbalancing} are as follows \cite{understanding-linux-kernel-3rd} \cite{linux-tovalds}: 

\begin{itemize}
\item
Rebalance\_tick: update the average load of the run-queue.
\item
Load\_balance: inspect the degree of load imbalance of the scheduling domain \cite{new-sched-domain-for-multicore}.
\item
Find\_busiest\_group: analyze the load of groups within the scheduling domain.
\item
Find\_busiest\_queue: search for the busiest CPU within the found group.
\item
Move\_tasks: migrate tasks from the source run-queue to the target run-queue in other CPU.
\item
Dequeue\_tasks: remove tasks from the external run-queue.
\item
Enqueue\_tasks: add tasks into a particular CPU.
\item
Resched\_task: if the priority of moved tasks is higher than that of current running tasks, preempt the current task of a particular CPU.
\end{itemize}

At every tick, the \ident{scheduler_tick()} function calls \ident{rebalance_tick()} function to adjust the load of the run-queue that is assigned to each CPU. At this time, load-balancer uses \ident{this_cpu} index of local CPU, \ident{this_rq}, \ident{flag}, and \texttt{idle (SCHED\_IDLE, NOT\_IDLE)} to make a decision. The \ident{rebalance_tick()} function determines the number of tasks that exist in the run-queue. It updates the average load of the run-queue by accessing \ident{nr_running} of the run-queue descriptor and \ident{cpu_load} field for all domains from the default domain to the domain of the upper layer. If the load imbalance is found, the SMP scheduler starts the procedure to balance the load of the scheduling domain by calling \ident{load_balance()} function. 

It is determined by \ident{idle} value in the \ident{sched_domain} descriptor and other parameters how frequently load-balancing happens. If \ident{idle} value is \ident{SCHED_IDLE}, meaning that the run-queue is empty, \ident{rebalance_tick()} function frequently calls \ident{load_balance()} function. On the contrary, if \ident{idle} value is \ident{NOT_IDLE}, the run-queue is not empty, and \ident{rebalance_tick()} function delays calling \ident{load_balance()} function. For example, if the number of running tasks in the run-queue increases, the SMP scheduler inspects whether the load-balancing time \cite{load-balancing-in-parallel} of the scheduling domain belonging to physical CPU needs to be changed from 10 milliseconds to 100 milliseconds.

When \ident{load_balance()} function moves tasks from the busiest group to the run-queue of other CPU, it calculates whether Linux can reduce the load imbalance of the scheduling domain. If \ident{load_balance()} function can reduce the load imbalance of the scheduling domain as a result of the calculation, this function gets parameter information like \ident{this_cpu}, \ident{this_rq}, \ident{sd}, and \ident{idle}, and acquires spin-lock called \ident{this_rq->lock} for synchronization. Then, \ident{load_balance()} function returns \ident{sched_group} descriptor address of the busiest group to the caller after analyzing the load of the group in the scheduling domain by calling \ident{find_busiest_group()} function. At this time, \ident{load_balance()} function returns the information of tasks to the caller to move the tasks into the run-queue of local CPU for the load-balancing of scheduling domain.

The kernel moves the selected tasks from the busiest run-queue to \ident{this_rq} of another CPU. After turning on the flag, it wakes up \ident{migration/*} kernel thread. The migration thread scans the hierarchical scheduling domain from the base domain of the busiest run-queue to the top in order to find the most idle CPU. If it finds relatively idle CPU, it moves one of the tasks in the busiest run-queue to the run-queue of relatively idle CPU (calling \ident{move_tasks()} function). If a task migration is completed, kernel releases two previously held spin-locks, \ident{busiest->lock} and \ident{this_rq->lock}, and finally it finishes the task migration.

\ident{dequeue_task()} function removes a particular task in the run-queue of other CPU. Then, \ident{enqueue_task()} function adds a particular task into the run-queue of local CPU. At this time, if the priority of the moved task is higher than the current task, the moved task will preempt the current task by calling \ident{resched_task()} function to gain the ownership of CPU scheduling. 

As we described above, the goal of the load-balancing is to equally utilize each CPU \cite{adjustable-process-scheduling}, and the load-balancing is performed after periodically checking whether the load of CPUs is fair. The load-balancing overhead is controlled by adjusting frequency of load-balancing operation, \ident{load_balance()} function, according to the number of running tasks in the run-queue of CPU. However, since it always performs load-balancing whenever a load imbalance is found, there is unnecessary load-balancing which does not help to improve overall system performance. 

In multicore embedded systems running many user applications at the same time, load imbalance will occur frequently. In general, more CPU load leads to more frequent task migration, and thus, incurs higher cost. The cost can be broken down into direct, indirect, and latency costs as follows:
\begin{enumerate}
\item
Direct cost: the load-balancing cost by checking the load imbalance of CPUs for utilization and scalability in the multicore system
\item
Indirect cost: cache invalidation and power consumption
   \begin{enumerate}
   \item cache invalidation cost by task migration among the CPUs
   \item power consumption by executing more instructions according to aggressive load-balancing
   \end{enumerate}
\item
Latency cost: scheduling latency and longer non-preemptible period
   \begin{enumerate}
   \item scheduling latency of the low priority task because the migration thread moves a number of tasks to another CPU \cite{rt-linux-yodaiken}
   \item longer non-preemptible period by holding the double-locking for task migration
   \end{enumerate}
\end{enumerate}

We propose our \emph{operation zone based load-balancer} in the next section to solve those problems.

% Article: 3. Operation zone based load-balancer.

\section{Operation zone based load-balancer}

In this section, we propose a novel load-balancing scheduler called \emph{operation zone based load-balancer} which flexibly migrates tasks for load-balancing based on \emph{load-balancing operation zone} mechanism which is designed to avoid too frequent unnecessary load-balancing. We can minimize the cost of the load-balancing operation on multicore systems while maintaining overall CPU utilization balanced. 

\begin{figure}  
\centering
\includegraphics[width=1.0\columnwidth]{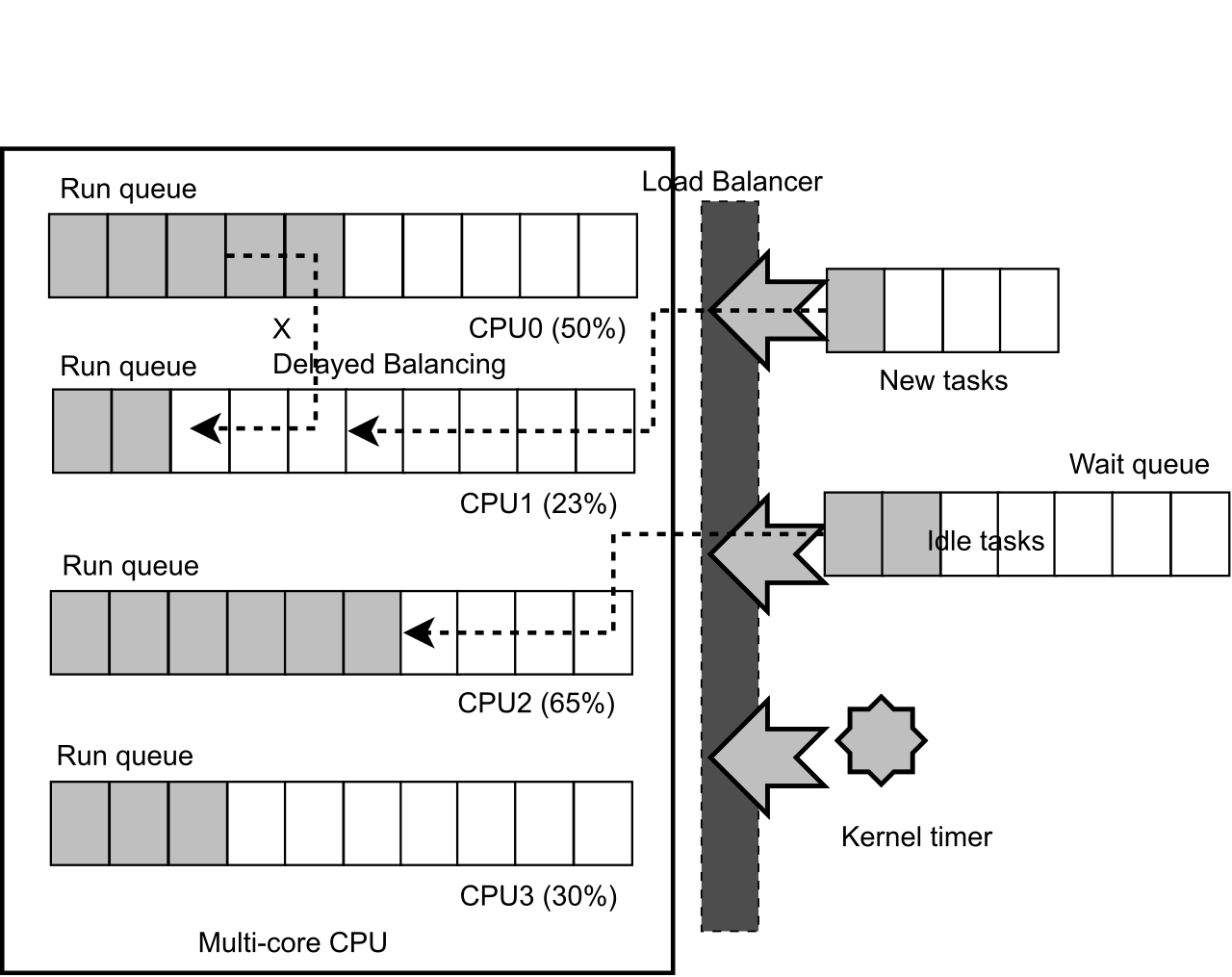}
\caption{Flexible task migration for low latency}
\label{lim-fig-flexible-migration-operation}
\end{figure}   

The existing load-balancer described in the previous section regularly checks whether load-balancing is needed or not. On the contrary, our approach checks only when the status of tasks can be changed. As illustrated in Figure~\ref{lim-fig-flexible-migration-operation}, \emph{operation zone based load-balancer} checks whether the task load-balancing is needed in the following three cases:
\begin{itemize}
\item
A task is newly created by the scheduler. 
\item
An idle task wakes up for scheduling. 
\item
A running task belongs to the busiest scheduling group. 
\end{itemize}

The key idea of our approach is that it defers load-balancing when the current utilization of each CPU is not seriously imbalanced. By avoiding frequent unnecessary task migration, we can minimize heavy double-lock overhead and reduce power consumption of a battery backed embedded device. In addition, it controls the worst-case scenario: one CPU load exceeds 100\% even though other CPUs are not fully utilized. For example, when a task in \ident{idle}, \ident{newidle}, or \ident{noactive} state is rescheduled, we can make the case that does not execute \ident{load_balance()} routine. 

\subsection{Load-balancing operation zone}
\begin{figure}    
\centering
\includegraphics[width=1.0\columnwidth]{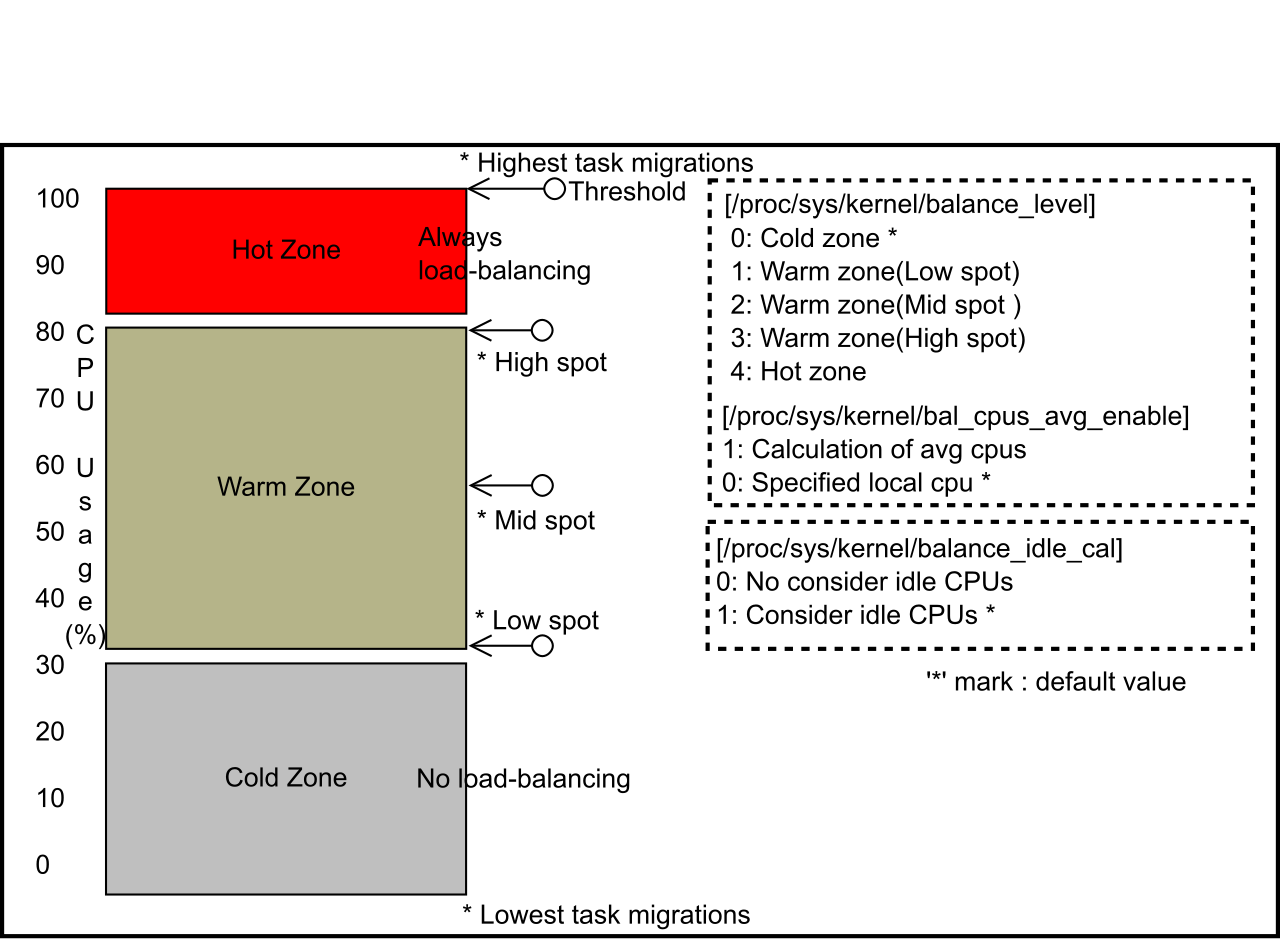}
\caption{Load-balancing operation zone}
\label{lim-fig-util-zone-based-lb}
\end{figure}         

Our \emph{operation zone based load-balancer} provides \emph{load-balancing operation zone} policy that can be configured to the needs of the system. As illustrated in Figure~\ref{lim-fig-util-zone-based-lb}, it provides three multicore load-balancing policies based on the CPU utilization. The \emph{cold zone} policy loosely performs load-balancing operation; it is adequate when the CPU utilization of most tasks is low. 

On the contrary, the \emph{hot zone} policy performs load-balancing operation very actively, and it is proper under high CPU utilization. The \emph{warm zone} policy takes the middle between \emph{cold zone} and \emph{hot zone}. 

Load-balancing under the \emph{warm zone} policy is not trivial because CPU utilization in \emph{warm zone} tends to fluctuate continuously. To cope with such fluctuations, \emph{warm zone} is again classified into three spots---high, mid, and low---and our approach adjusts scores based on weighted values to further prevent unnecessary task migration caused by the fluctuation. We provide {\tt /proc} interfaces for a system administrator to configure the policy either statically or dynamically. From our experience, we recommend that a system administrator configures the policy statically because of system complexity.  

\subsubsection{Cold zone}
In a multicore system configured with the \emph{cold zone} policy, our operation zone based load-balancing scheduler does not perform any load-balancing if the CPU utilization is in \emph{cold zone}, 0\textasciitilde 30\%. Since there is no task migration in \emph{cold zone}, a task can keep using the currently assigned CPU. Kernel performs the load-balancing only when the CPU utilization exceeds \emph{cold zone}. 

This policy is adequate where the CPU utilization of a device tends to be low except for some special cases. It also helps to extend battery life in battery backed devices. 

\subsubsection{Hot zone}
Task migration in the \emph{hot zone} policy is opposite to that in the \emph{cold zone} policy. If the CPU utilization is in \emph{hot zone}, 80\textasciitilde 100\%, kernel starts to perform load-balancing. Otherwise, kernel does not execute the procedure of load-balancing at all. 

Under the \emph{hot zone} policy, kernel defers load-balancing until the CPU utilization reaches \emph{hot zone}, and thus, we can avoid many task migrations. This approach brings innovative results in the multicore-based system for the real-time critical system although the system throughput is lost.

\subsubsection{Warm zone}
In case of the \emph{warm zone} policy, a system administrator chooses one of the following three spots to minimize the costs of the load-balancing operation for tasks whose CPU usage is very active.  

\begin{itemize}
\item
High spot (80\%): This spot has the highest CPU usage in the \emph{warm zone} policy. The task of \emph{high spot} cannot go up any more in the \emph{warm zone} policy
\item
Low spot (30\%): This spot has the lowest CPU usage in the \emph{warm zone} policy. The task of \emph{low spot} cannot go down any more in the \emph{warm zone} policy.
\item
Mid spot (50\%): This spot is in between high spot and low spot. The weight-based dynamic score adjustment scheme is used to cope with fluctuations of CPU utilization. 
\end{itemize}

The spot performs the role of controlling the CPU usage of tasks that can be changed according to weight score. In the \emph{warm zone} policy system, weight-based scores are applied to tasks according to the period of the CPU usage ratio based on \emph{low spot}, \emph{mid spot} and \emph{high spot}. The three spots are detailed for controlling active tasks by users. The CPU usages of tasks have penalty points or bonus points according to the weight scores.

\begin{figure}      
\centering
\includegraphics[width=1.0\columnwidth]{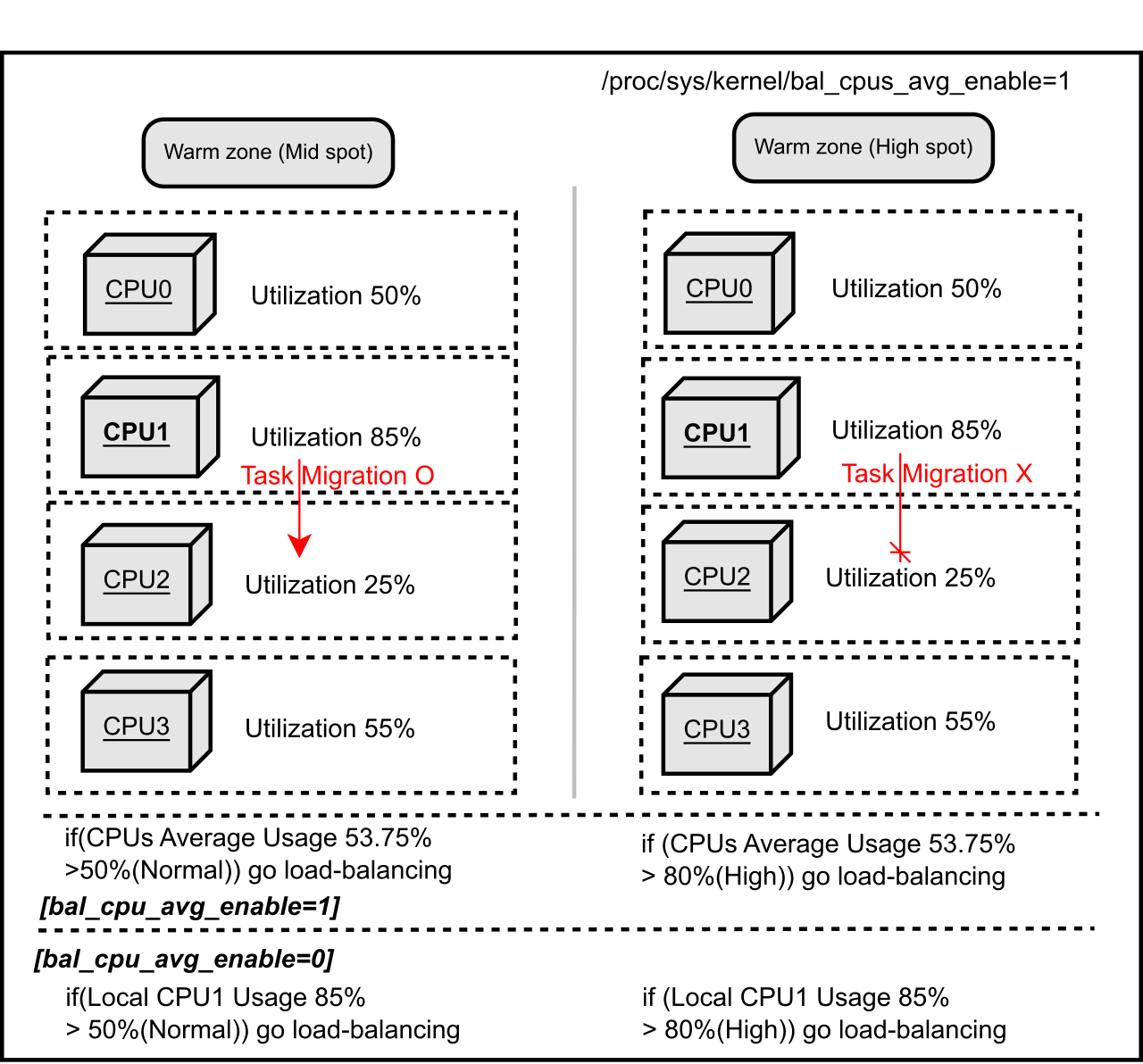}
\caption{Task migration example in warm zone policy}
\label{lim-fig-task-migration-example}
\end{figure}         

Although the score of task can increase or decrease, these tasks cannot exceed the maximum value, \emph{high spot}, and go below the minimum value, \emph{low spot}. If CPU usage of a task is higher than that of the configured spot in the \emph{warm zone} policy, kernel performs load-balancing through task migration. Otherwise, kernel does not execute any load-balancing operation. 

For example, we should consider that the CPU utilization of quad-core systems is 50\%, 85\%, 25\% and 55\% respectively from CPU0 to CPU3 as Figure~\ref{lim-fig-task-migration-example}. If the system is configured in \emph{mid spot} of the \emph{warm zone} policy, the load-balancer starts operations when the average usage of CPU is over 50\%. Kernel moves one of the running tasks of run-queue in CPU1 with the highest utilization to the run-queue of CPU2 with the lowest utilization. 

In case of \emph{high spot} and the \emph{warm zone} policy, the load-balancer starts the operations when the average usage of CPU is over 80\%. Tasks that are lower than the CPU usage of the \emph{warm zone} area is not migrated into another CPU according to migration thread. Figure~\ref{lim-fig-task-migration-example} depicts the example of load-balancing operations on the \emph{warm zone} policy. 

\begin{figure}  
\centering
\includegraphics[width=1.0\columnwidth]{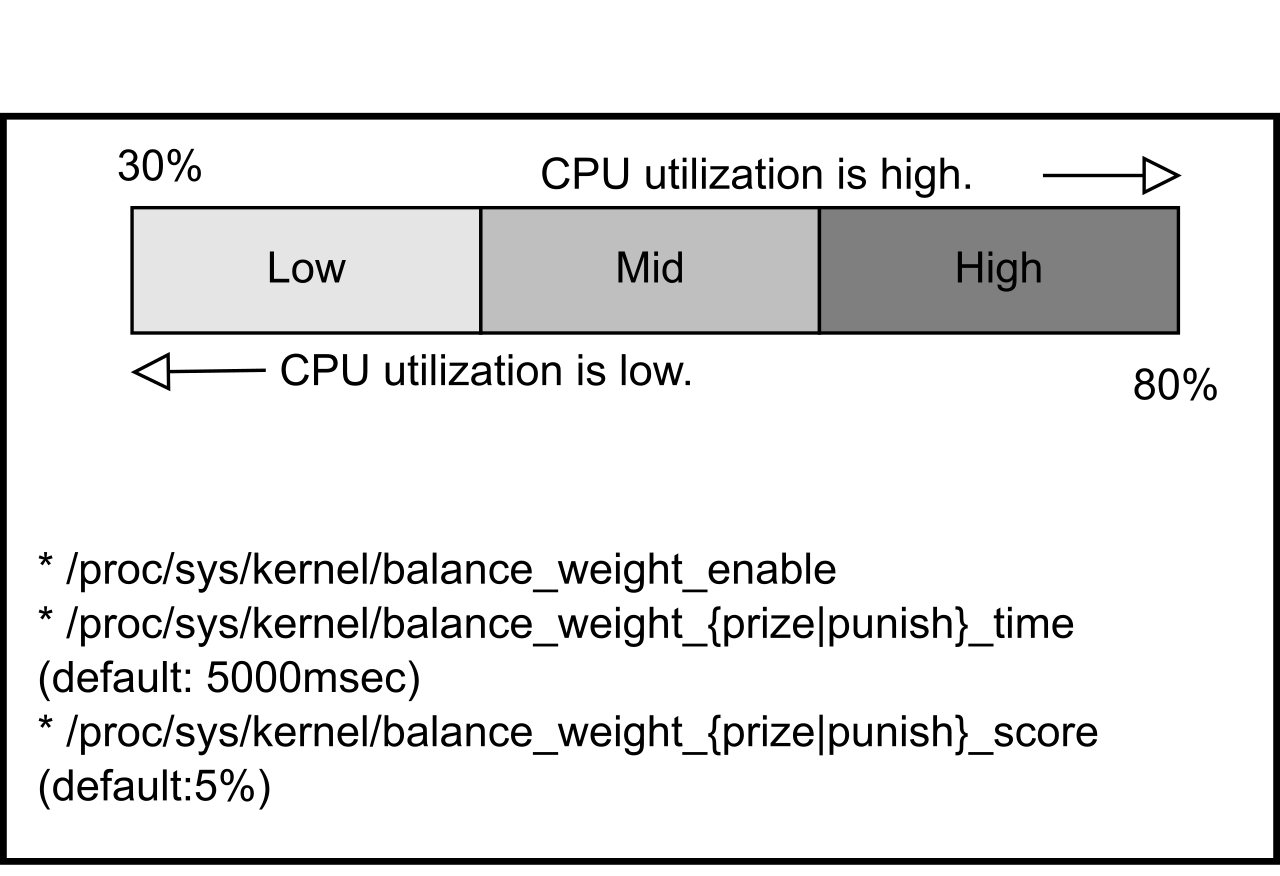}
\caption{Weight-based score management}
\label{lim-fig-weight-based-score}
\end{figure}

Figure~\ref{lim-fig-weight-based-score} shows weight-based load score management for the \emph{warm zone} policy system. When the usage period of CPU is longer than the specified time, five seconds by default, kernel manages bonus points and penalty points to give relative scores to the task that utilizes CPU resources continually and actively. Also, kernel operates the load weight-based \emph{warm zone} policy to support the possibility that the task can use the existing CPU continually. 

At this time, tasks that reach the level of the \emph{high spot}, stay in the \emph{warm zone} range although the usage period of CPU is very high. Through these methods, kernel keeps the border of the \emph{warm zone} policy without moving a task to the \emph{hot zone} area. 

If a task maintains the high value of the usage of CPU more than five seconds as the default policy based on \code{/proc/sys/kernel/balance_weight_{prize|punish}_time}, kernel gives the task CPU usage score of -5 which means that CPU utilization is lower. At this point, the CPU usage information of the five seconds period is calculated by the scheduling element of a task via \emph{proc file system}. We assigned the five seconds by default via our experimental experience. This value can be changed by using \code{/proc/sys/kernel/balance_weight_{prize|punish}_time} by the system administrator to support various embedded devices. 

In contrast, if a task consumes the CPU usage of a spot shorter than five seconds, kernel gives the task CPU usage score of +5 which means that CPU utilization is higher. The task CPU usage score of +5 elevates the load-balancing possibility of tasks. Conversely, the task CPU usage score of -5 aims to bring down the load-balancing possibility of tasks. 

The value of the \emph{warm zone} policy is static, which means it is determined by a system administrator without dynamic adjustment. Therefore, we need to identify active tasks that consume the usage of CPUs dynamically. The load weight-based score management method calculates a task's usage in order that kernel can consider the characteristics of these tasks. This mechanism helps the multicore-based system manage the efficient load-balancing operation for tasks that have either high CPU usage or low CPU usage.

\subsection{Calculating CPU utilization}
In our approach, the CPU utilization plays an important role in determining to perform load-balancing. In measuring CPU utilization, our approach provides two ways: calculating CPU utilization for each CPU and averaging CPU utilization of all CPUs. A system administrator also can change behaviors through \ident{proc} interface, \code{/proc/sys/kernel/balance_cpus_avg_enable}. By default, kernel executes task migration depending on the usage ratio of each CPU. 

If a system administrator selects \code{/proc/system/kernel/balance_cpus_avg_enable=1} parameter for their system, kernel executes task migration depending on the average usage of CPUs. 

The method to compare load-balancing by using the average usage of CPUs, helps to affinitize the existing CPU as efficiently as possible for some systems. The system needs the characteristics of CPU affinity \cite{cache-affinity-on-multicore} although the usage of a specific CPU is higher than the value of the \emph{warm zone} policy, e.g.\ CPU-intensive single-threaded application in the most idle systems.

% Article: 4. evaluation.
\section{Evaluation}

\subsection{Evaluation scenario}
\begin{figure}   
\centering
\includegraphics[width=1.0\columnwidth]{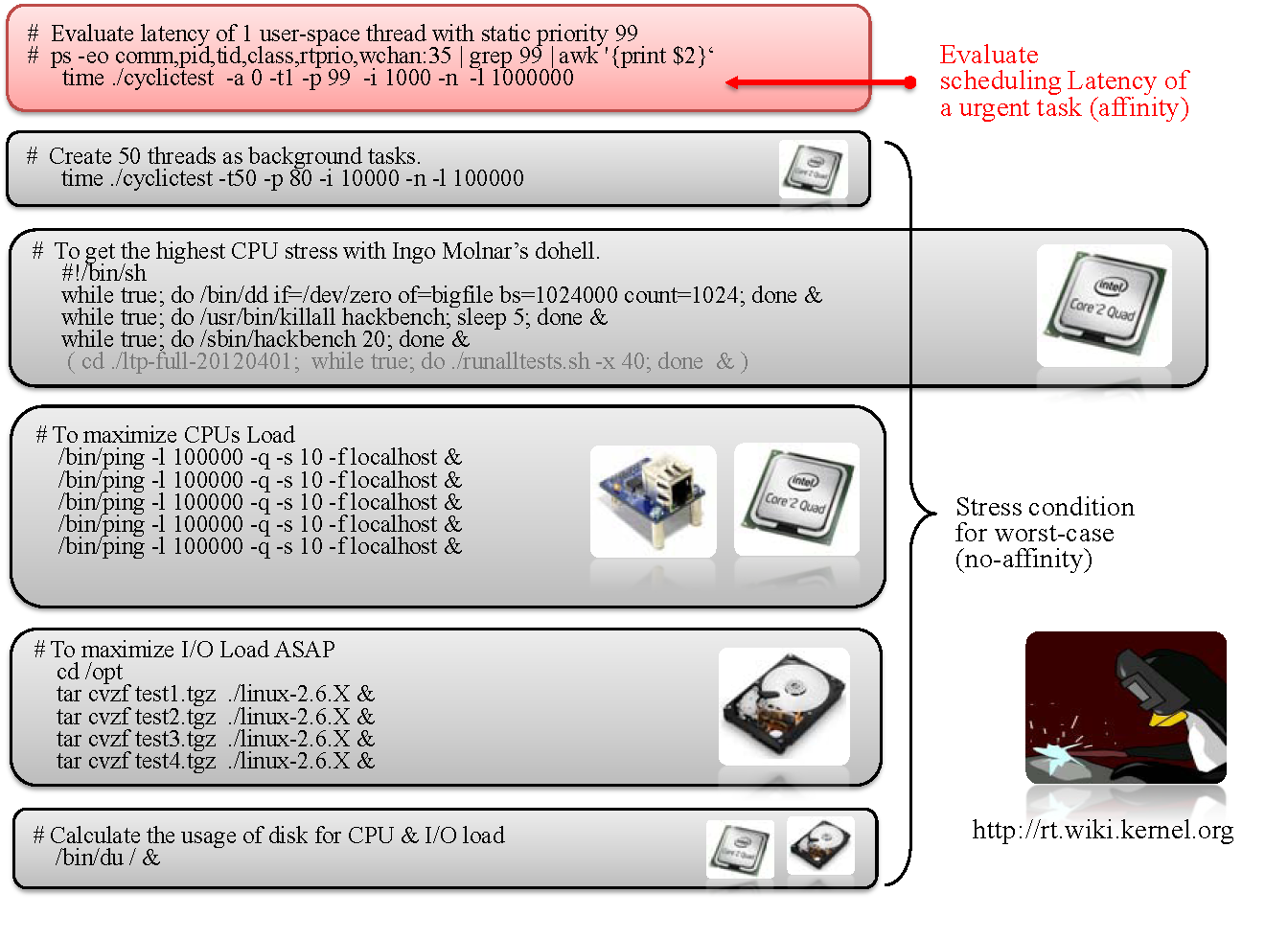}
\caption{Evaluation scenario to measure scheduling latency}
\label{lim-fig-evaluation-scenario}
\end{figure}

Figure \ref{lim-fig-evaluation-scenario} shows our evaluation scenario to measure the real-time characteristics of running tasks in multicore based embedded systems. In this experiment, we measured how scheduling latency of an urgent task would be reduced under very high CPU load, network stress, and disk I/O. 

To measure scheduling latency, we used \emph{cyclictest} utility of \emph{rt-test} package \cite{rt-tests} which is mainly used to measure real-time characteristics of \emph{Redhat Enterprise Linux (RHEL)} and real-time Linux. All experiments are performed in \ident{Linux 2.6.32} on \ident{Intel Quad core Q9400}. 

\subsection{Experimental result}

\begin{figure}    
\centering
\includegraphics[width=1.0\columnwidth]{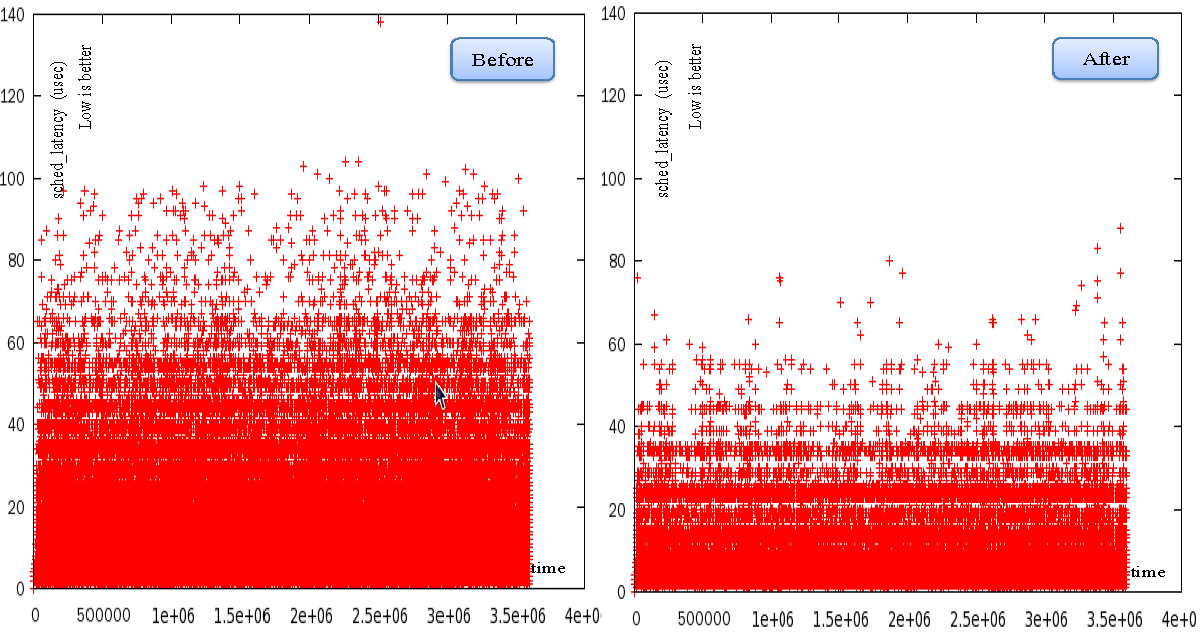}
\caption{Comparison of scheduling latency distribution}
\label{lim-fig-evaluation-result}
\end{figure}

In Figure \ref{lim-fig-evaluation-result}, we compared the scheduling latency distribution between the existing approach (before) and our proposed approach (after). 

Our approach is configured to use \emph{warm zone - high spot} policy. Under heavy background stress reaching to the worst load to the Quad-core system, we measured the scheduling latency of our test thread which repeatedly sleeps and wakes up. Our test thread is pinned to a particular CPU core by setting CPU affinity \cite{affinity-scheduling-on-smp-patent} and is configured as the FIFO policy with priority 99 to gain the best priority. 

In the figure, X-axis is the time from the test start, and Y-axis is the scheduling latency in microseconds from when it tries to wake up for rescheduling after a specified sleep time. As Figure \ref{lim-fig-evaluation-result} shows, the scheduling latency of our test thread is reduced more than two times: from 72 microseconds to 31 microseconds on average. 

In order to further understand why our approach reduces scheduling latency more than two times, we traced the caller/callee relationship of all kernel function during the experiment by using Linux internal function tracer, \emph{ftrace} \cite{ftrace}. 

The analysis of the collected traces confirms three: first, the scheduling latency of a task can be delayed when migration of other task happens. Second, when task migration happens, non-preemptible periods are increased for acquiring double-locking. Finally, our approach can reduce average scheduling latency of tasks by effectively removing vital costs caused by the load-balancing of the multicore system. 

In summary, since the migration thread is a real-time task with the highest priority, acquiring double-locking and performing task migration, the scheduling of the other tasks can be delayed. Since load imbalance frequently happens under a heavily loaded system with many concurrent tasks, the existing very fair load balancer incurs large overhead, and our approach can reduce such overhead effectively. 

Our \emph{operation zone based load-balancer} performs load-balancing based on CPU usage with lower overhead while avoiding overloading to a particular CPU that can increase scheduling latency. Moreover, since our approach is implemented only in the operating system, no modifications of user applications are required.

% Article: 5. further work.
\section{Further work}

In this paper, we proposed an operation zone based load-balancing mechanism which reduces scheduling latency. Even though it reduces scheduling latency, it does not guarantee deadline for real-time systems where the worst case is most critical. In order to extend our approach to the real-time tasks, we are considering a hybrid approach with the physical CPU shielding technique \cite{shield-processor-response} which dedicates a CPU core for a real-time task. We expect that such approach can improve real-time characteristics of a CPU intensive real-time task. 

Another important aspect especially in embedded systems is power consumption. In order to keep longer battery life, embedded devices dynamically turn on and off CPU cores. To further reduce power consumption, we will extend our load-balancing mechanism considering CPU on-line and off-line status. 

We experimented with scheduling latency to enhance the user responsiveness on the multicore-based embedded system in this paper. We have to evaluate various scenarios such as direct cost, indirect cost, and latency cost to use our load-balancer as a next generation SMP scheduler.

% Article: 6. conclusion.
\section{Conclusions}

We proposed a novel \emph{operation zone based load-balancing technique} for multicore embedded systems. It minimized task scheduling latency induced by the load-balancing operation. Our experimental results using the \emph{cyclictest} utility \cite{rt-tests} showed that it reduced scheduling latency and accordingly, users' waiting time.
 
Since our approach is purely kernel-level, there is no need to modify user-space libraries and applications. Although the vanilla Linux kernel makes every effort to keep the CPU usage among cores equal, our proposed \emph{operation zone based load-balancer} schedules tasks by considering the CPU usage level to settle the load imbalance. 

Our design reduces the non-preemptible intervals that require double-locking for task migration among the CPUs, and the minimized non-preemptible intervals contribute to improving the software real-time characteristics of tasks on the multicore embedded systems. 

Our scheduler determines task migration in a flexible way based on the \emph{load-balancing operation zone}. It limits the excess of 100\% usage of a particular CPU and suppresses the task migration to reduce high overhead for task migration, cache invalidation, and high synchronization cost. It reduces power consumption and scheduling latency in multicore embedded systems, and thus, we expect that customers can use devices more interactively for longer time.

% Article: 7. acknowledgement.
\section{Acknowledgments}

We thank Joongjin Kook, Minkoo Seo, and Hyoyoung Kim for their feedback and comments, which were very helpful to improve the content and presentation of this paper. This work was supported by the IT R\&D program of MKE/KEIT [10041244, SmartTV 2.0 Software Platform].

\begin{flushleft}
\bibliography{main}
\bibliographystyle{plain}
\end{flushleft}

% Some bibliography styles
%
%1: ieeetr
%2: unsrt
%3: IEEE
%4: ama
%5: cj
%6: nar
%7: nature
%8: phjcp
%9: is-unsrt 	
%10: plain
%11: abbrv
%12: acm
%13: siam
%14: jbact
%15: amsplain
%16: finplain
%17: IEEEannot
%18: is-abbrv 	
%19: is-plain
%20: annotation
%21: plainyr
%22: decsci
%23: jtbnew
%24: neuron
%25: cell
%26: jas99
%27: abbrvnat 	
%28: ametsoc
%29: apalike
%30: jqt1999
%31: plainnat
%32: jtb
%33: humanbio
%34: these
%35: chicagoa
%36: development 	
%37: unsrtnat
%38: amsalpha
%39: alpha
%40: annotate
%41: is-alpha
%42: wmaainf
%43: alphanum
%44: apasoft 
%99: manual method (directly)

\end{document}